\documentclass[11pt]{article}
\usepackage{a4wide}

\usepackage{amsmath, amssymb, amsthm, amsfonts, dsfont, bbm,calc}
\usepackage{enumerate}

\def\phi{\varphi}
\def\et{\,\&\,}

\newcommand{\mathrmL}{{\mathchoice{\mbox{\rm\L}}{\mbox{\rm\L}}{\mbox{\rm\scriptsize\L}}{\mbox{\rm\tiny\L}}}}
\newtheorem{theorem}{Theorem}[section]
\newtheorem{definition}[theorem]{Definition}
\newtheorem{lemma}[theorem]{Lemma}

\newtheorem{corollary}[theorem]{Corollary}

\newtheorem{problem}[theorem]{Problem}

\newcommand{\logic}[1]{\ensuremath{\mathrm {#1}}}
\newcommand\FL[1]{\ensuremath{ \logic{{FL}}_\mathrm{#1} }}
\newcommand\FLew{\FL{ew}}

\newcommand\BL{\logic{BL}}

\newcommand\RPL{\logic{RPL}}
\newcommand\Lpred{\logic{\mathrmL}\forall}
\newcommand\RPLpred{\logic{RPL}\forall}
\newcommand\infL{\mathrmL_{\infty}}

\newcommand{\alg}[1]{ {\ensuremath{\mathcal {#1}}}}

\newcommand\standardL{\ensuremath{{[0,1]_\mathrmL}}}

\newcommand\finiteMV[1]{ \ensuremath{ \alg{L}_{#1} }}

\newcommand\SAT{\ensuremath{\mathrm{SAT}}}

\newcommand\DEF{\ensuremath{\mathrm{DEF}}}

\newcommand\ComplexityClass[1]{\ensuremath{\mathrm{#1}}}

\newcommand\NP{\ensuremath{\ComplexityClass{NP}}}
\newcommand\coNP{\ensuremath{\ComplexityClass{coNP}}}

\begin{document}

\title{Implicit definability of truth constants in {\L}ukasiewicz logic}
\author{Zuzana Hanikov\'a\\
Institute of Computer Science\\
Czech Academy of Sciences\\ 
182 07 Prague, Czech Republic\\
hanikova@cs.cas.cz}
\maketitle

\begin{abstract}
In the framework of propositional {\L}ukasiewicz logic, a suitable notion of implicit definability, 
tailored to the intended real-valued semantics and referring to the elements of its domain, is introduced.
Several variants of implicitly defining each of the rational elements in the standard semantics are explored,
and based on that, a faithful interpretation of theories in Rational Pavelka logic 
in theories in {\L}ukasiewicz logic is obtained. 
Some of these results were already presented in \cite{Hajek:1998} as technical statements. 
A connection to the lack of (deductive) Beth property in {\L}ukasiewicz logic is drawn. 
Moreover, while irrational elements of the standard semantics
are not implicitly definable by finitary means, 
a parallel development is possible for them in the infinitary {\L}ukasiewicz logic.
As an application of definability of the rationals, it is shown how 
computational complexity results for Rational Pavelka logic 
can be obtained from analogous results for {\L}ukasiewicz logic. 
The complexity of the definability notion itself is studied as well. 
Finally, we review the import of these results for the precision/vagueness discussion for fuzzy logic,
and for the general standing of truth constants in {\L}ukasiewicz logic.
\end{abstract}

\section{Introduction}
\label{section:intro}


This paper is a contribution to the study of truth constants in {\L}ukasiewicz logic. 
Previous works on the subject, often in a wider context of other t-norm based logics,
include \cite{Pavelka:Fuzzy,Hajek:1998,Hajek-Paris-Shepherdson:Pavelka,Esteva-GGN:AddingTruthConstants,Cintula:noteAxiomatizationsPavelka}.
Predominantly, this paper is about constants for rational elements of $[0,1]$,
but an analogous development 
for irrational constants in the infinitary variant of {\L}ukasiewicz logic is suggested.  
Since truth constants are native to propositional logic, the entire paper stays on the propositional level.

Two logics are juxtaposed in this paper: (infinite-valued) {\L}ukasiewicz logic $\mathrmL$ and 
Rational Pavelka logic RPL.
The former was introduced by Jan {\L}ukasiewicz, 
with later simplifications by himself and others (see \cite{Lukasiewicz-Tarski:Untersuchungen}). 
The latter, RPL, was introduced by H\'ajek in \cite{Hajek:1998} as a considerable simplification of earlier systems for reasoning with partially true statements
as envisaged by Goguen \cite{Goguen:LogicInexactConcepts} and proposed by Pavelka \cite{Pavelka:Fuzzy}. 
Many resources discuss one or both of these logics very accurately, but
two excellent sources are \cite{Hajek:1998} and \cite{Hajek-Paris-Shepherdson:Pavelka}.

In Section \ref{section:definability}, 
we start with providing a semantic notion of implicit definability, referring to the standard MV-algebra $\standardL$. 
This is in accord with the ontology of truth constants (rational or other) in {\L}ukasiewicz logic,
which arise from the real-valued semantics (intended, and presumably therefore called ``standard'') and permeate the syntax. 
Using the fact that
 that each rational is definable in $\standardL$ under this semantic definability notion, 
we faithfully interpret RPL in a particular theory over $\mathrmL$; this interpretation result
was first presented in \cite{Hajek:1998}, where a different evidence for implicit definability is provided.
Moreover, \cite[Theorem 2.6]{Hajek-Paris-Shepherdson:Pavelka} offers what is presumably the next best thing 
to rational constants being term definable: for each formula $\phi$ of $\RPL$ and positive integer $n$ divisible
by denominators of truth constants in $\phi$, there is an $\mathrmL$-formula $\psi$ s.t.~$\RPL\vdash\phi^n\equiv\psi$.
Along the way, we also discuss other methods of obtaining implicit definability of rationals; e.g., 
one can rely on the bookkeeping formulas, which do the job very adequately, or one can refer to formulas representing
certain McNaughton functions. 

Mimicking the finitary definition in 
(finitary) {\L}ukasiewicz logic,  a non-finitary definability notion for irrationals
in  infinitary {\L}ukasiewicz logic $\infL$ is outlined, 
with some analogous properties (e.g., under a theory containing the defining formulas,
valid bookkeeping statements for the irrationals can be derived in the logic).

Implicit definability is evocative of Beth property. We briefly consider this (somewhat broader) context,
and discuss how our technique provides another way of demonstrating the lack of (deductive) Beth property in
{\L}ukasiewicz logic,  proved in \cite{Hoogland:Thesis}. 
The status of RPL with respect to Beth property seems to be unknown.

We then weigh the message of this interpretability result, and exploit it somewhat. 
It seems to be often overlooked that
a considerable part of the metamathematics of {\L}ukasiewicz logic with constants is already 
addressed by (a special case of) the metamathematics of theories over {\L}ukasiewicz logic.
As an example of this phenomenon and an 
application of the interpretation result, in Section \ref{section:complexity} 
we inspect the result on complexity of propositional RPL,
that is, the recognition of theoremhood and finite consequence relation. Both are $\coNP$-complete;
we show that this result is implied by (the same) complexity results for $\mathrmL$. 
We also show that the notion of definability as such, taken as a decision problem, is hard.

Finally, apparently the defining formulas allow for speaking about arbitrary rational truth values, 
which brings to mind previous discussions of the artificial precision problem in fuzzy logic.
We briefly touch this topic in Section \ref{section:vagueness}. 

\section{Preliminaries}

This paper deals with propositional {\L}ukasiewicz logic and its algebraic semantics.
It makes no distinction between propositional formulas (in the language of {\L}ukasiewicz or other logics, possibly expanding the language) 
and terms of the language of the algebraic semantics;
thus \emph{formulas} and \emph{terms} are the same objects.
Similarly, we conflate propositional connectives with function symbols. 

Although {\L}ukasiewicz logic can be presented in several very succinct sets of basic connectives,
we make use of a variety of connectives: the language of $\FLew$, namely, the set $\{\cdot,\to,\land,\lor, \overline{0},\overline{1} \}$,
the negation $\neg$, the strong disjunction $\oplus$, and the equivalence $\equiv$.
The expression $x^n$ stands for $\underbrace{x\cdot x\cdot \dots x}_{n \text{ times}}$ 
and $nx$ stands for $\underbrace{x\oplus x\oplus \dots x}_{n \text{ times}}$.

{\L}ukasiewicz logic is amply presented and discussed in literature 
(see \cite{Cignoli-Ottaviano-Mundici:AlgebraicFoundations,DiNola-Leustean:HandbookMValgebras,Mundici:Advanced}); 
this paper takes for granted the reader's  familiarity with the basics. 
A general semantics of {\L}ukasiewicz logic is given by the variety of MV-algebras.
The \emph{standard} MV-algebra $\standardL$ is given by the {\L}ukasiewicz t-norm 
on the domain of the real unit interval $[0,1]$. 
While $\mathrmL$ is strongly complete w.r.t.~MV-algebras, it only 
enjoys finite strong standard completeness, but not strong standard completeness (see \cite{Hajek:1998, Cintula-EGGMN:DistinguishedSemantics}).
The finite MV-chain with $n+1$ elements is denoted $\finiteMV{n+1}$.

Rational Pavelka logic $\RPL$ (see \cite{Hajek:1998}) expands the language of {\L}ukasiewicz logic with constants for rationals in $[0,1]$.
Extending the notational convention for the two constants $\overline{0}$ and $\overline{1}$, 
we use horizontal bars to distinguish the constant $\overline{r}$ from its intended interpretation $r$ (this convention extends also to irrationals used here).\footnote{We also use a horizontal bar to denote 
complement of a set. A vector of variables is denoted with a tilde (such as $\tilde{x}$).}
The intended interpretation, i.e., $\overline{r}$ being interpreted by $r$ for each $r\in \mathds{Q}$, is referred to as
the \emph{canonical} interpretation of constants, to distinguish it from other interpretations, given by algebras that contain an isomorphic copy of (a subalgebra of) the standard MV-algebra on the rationals.

A \emph{bookkeeping formula} (see \cite{Hajek:1998}) is a formula recording the behaviour of an algebraic operation on elements of the domain 
interpreting the constants, predominantly, the rationals in $[0,1]$; this domain is closed under the operations of $\standardL$.
For example,  $\overline{6/13} \to \overline{5/13} \equiv \overline{12/13}$ is a bookkeeping formula.
It is quite usual to include \emph{all} such bookkeeping formulas that are valid under the canonical interpretation in $\standardL$
as \emph{axioms} of a logic featuring constants; the bookkeeping axioms are then the only new axioms that are added. 
This is indeed the case of $\RPL$ as an axiomatic expansion of $\mathrmL$.

We do not consider a general semantics for $\RPL$,  only the standard one, whose {\L}-reduct is the algebra $\standardL$.
By slight abuse of language, we use the notation $\standardL$ also for the $\RPL$-algebra that expands the standard MV-algebra
with a canonical interpretation of each rational constant, or even for an algebra that further expands the language with
constants whose intended interpretation is irrational (Subsection \ref{subsection:irrationals}).

$\RPL$ enjoys finite strong standard canonical completeness (see \cite{Esteva-GGN:AddingTruthConstants}): 
that is, for a finite $T,\varphi$ in the language of $\RPL$,
we have $T\vdash_{\RPL}\phi$ if and only if $T$ entails $\phi$ in $\standardL$ 
under the canonical interpretation of constants.\footnote{In fact, the canonical interpretation of constants 
is the only one that consistently expands $\standardL$ under the usual bookkeeping axioms; see Section \ref{section:definability}.}

Finite strong canonical completeness of $\RPL$ (and finite strong completeness of $\mathrmL$) entails 
a conservativity result for $\RPL$ over $\mathrmL$ for propositional case; 
the conservativity proof for the two first-order logics is more laborious (see \cite{Hajek-Paris-Shepherdson:Pavelka}).
Although the proof is based on \emph{finite} strong standard completeness, we do have that if 
$T$ is infinite and $T,\phi$ are without constants, then 
 $T\vdash_{\RPL}\phi$ if and only if $T\vdash_\mathrmL \phi$.

\section{Definability}
\label{section:definability}

A truth constant is specified within the language of a propositional calculus: 
it is a nullary connective, its behaviour determined by the axioms and rules of the calculus.
Providing a semantics for the calculus involves suggesting an interpretation of its terms,
including the constants.

Nevertheless, in addition to constants specified by the language, 
some other terms in the language may behave as constants too. 
As a ready example, many terms of either classical or fuzzy propositional logic behave as 
the constants $\overline{0}$ and $\overline{1}$: 
we refer to them as \emph{contradictions} and \emph{tautologies} respectively.
Here ``behave as'' can have either of two meanings: the constant $\overline{0}$ 
is in the language anyway and the term in question
is provably equivalent to it, or it is not in the language and one relies on the a semantics
where the term in question always evaluates to the intended evaluation of the 
(possibly absent) constant.\footnote{The discourse in mathematical fuzzy logic so far 
confirms that fuzzy logics are semantics-based; accordingly, 
\emph{constants} are tied to the intended (real-valued) semantics, standing for the rational or the real numbers thereof.
Indeed, where all rational constants from the interval $[0,1]$ are present, 
one may claim that semantics has leaked into syntax in a substantial way, 
in particular, such a process narrows down the range of algebraic interpretations of the calculus noticeably.}
In either case, one would speak about term definability. We return to this kind of definability in Section \ref{subsection:beth}.
Recalling that the set of contradictory terms and the set of tautologous terms 
of classical propositional are both $\coNP$ complete, 
the simple example above also hints that it is nontrivial to determine 
if a given term behaves as a particular constant.

\begin{definition}
Let $a\in [0,1]$. The value $a$ is term-definable in $\standardL$  if and only if 
there is an MV-term $\phi$ such that $v(\phi) =a$ for each valuation $v$ in $\standardL$.
\end{definition}

In the standard MV-algebra $\standardL$,  
no constants  beyond the classical ones
are term-definable: constant non-integer functions are not McNaughton functions (\cite{McNaughton:FunctionalRep}).
So the decision problem ``Does $\phi$ term-define $x$?'' 
is limited to $x\in\{0,1\}$; we are asking about unsatisfiable or tautologous terms
in the real-valued semantics of propositional {\L}ukasiewicz logic.

\subsection{Defining the rationals}

This section aims to expose a phenomenon that is one step more nuanced.
Namely, some variables may behave as constants in the framework of a particular theory:
 in each model of the theory, the value of such variables is fixed.
 We will discuss (propositional) \emph{implicit definability}.
In fact we shall present two different concepts: implicit definability by a term (i.e., a finite theory)
and a variant thereof where the defining theory is necessarily infinite and the logic considered is not finitary. 
The focus of this paper is on the former concept, which employs the more frequently considered \emph{finitary} {\L}ukasiewicz logic $\mathrmL$.

\begin{definition}
\label{def_definability}
Let $\alg{A}$ be an MV-algebra, $a$ an element of its domain, $\phi(x_1,\dots,x_n)$ an MV-term, and $1\leq i\leq n$. 
The term $\phi$ implicitly defines the element $a$ in variable $x_i$ in $\alg{A}$  
if and only if 
\begin{itemize}
\item $\phi$ is satisfiable in $\alg{A}$, and
\item $v(\phi)=1^\alg{A}$ implies $v(x_i)=a$ for each $\alg{A}$-valuation $v$.
\end{itemize}
An element $a\in \alg{A}$ is definable in $\alg{A}$ if and only if there is an MV-term that defines it there. 
\end{definition}




It is not difficult to come up with a suitable theory that implicitly defines
 rationals in $\standardL$. Indeed a theory $T$ in variables indexed by the rationals (say, $x_{m/n}$ for $m,n$ coprime), 
and with just the usual bookkeeping formulas for these variables, obtained from the corresponding MV-operations
on the indices of the variables (e.g.,  $x_{6/13} \to x_{5/13} \equiv x_{12/13}$), as the axioms of $T$, makes 
all the variables $x_{m/n}$ behave just as the constant $\overline{m/n}$ would behave in RPL.
$T$ is infinite, so it does not follow the requirement of implicit definability (of each of the rationals) by a term;
however, one quickly observes that to implicitly define $m/n$, only constants with denominators $n$ are needed.
This allows for a much narrower theory $T$ 
that is finite and has a unique interpretation (namely, the canonical one)  in $\standardL$.
We shall return to this claim at the end of Subsection \ref{subsection:theory} 
by providing a sketchy example, and we also discuss a redundancy issue there.

 Another way to implicitly define a rational $a$ is to rely on McNaughton functions:
 there is a plethora of one-variable McNaughton functions that attain the value 1 on the singleton $\{a\}$,
 each of them corresponds to an MV-term, and each such term implicitly defines $a$.

One construction using yet different term to achieve implicit definability was already given in \cite{Hajek:1998}. 
The book also presents a faithful interpretation of theories in RPL in theories in {\L}, which we reproduce below. 
We use the various versions of proving implicit definability for a general point to be made in a subsequent section.

We offer a simple variant of the definability axioms.
They are only marginally different from bookkeeping,
one of the differences being that $1/n$ is defined by a single term without additional variables.

\subsection{A theory of constants}
\label{subsection:theory}

From now on, \emph{definability} and \emph{implicit definability} may be conflated to the latter, for the sake of brevity. 

The following technical lemma is a useful tool for MV-algebras 
(cf.~ \cite{Torrens:CyclicElements,Gispert:UniversalClassesMV,Hanikova-Savicky:SAT}).

\begin{lemma}
\label{lemma_rat_l1}
Let $\alg{A}$ be an MV-chain and $n\geq 2$ an integer. 
The equation $x = (\neg x)^{n-1}$ has a solution in $\alg{A}$ 
if and only if $\alg{A}$ has a subalgebra isomorphic to $\mathrmL_{n+1}$.
If the solution exists, it is unique: the smallest nonzero element of $\mathrmL_{n+1}$.
\end{lemma}

\begin{corollary}
\label{lemma1n}
Let $n\geq 2$. The formula $x \equiv (\neg x)^{n-1}$ defines $1/n$ in $\standardL$. 
\end{corollary}

\begin{corollary}
\label{corollarymn}
Any rational number in the unit interval is definable  in $\standardL$. 
\end{corollary}

\begin{proof}
In $\standardL$,
the formulas $x\equiv \overline{0}$ and $x\equiv \overline{1}$ define rationals $0$ and $1$. Moreover $1/n$ is definable for $n\geq 2$ by
a formula in one variable.
The two-variable formula $\phi(x,y)$, defined as $(y\equiv(\neg y)^{n-1}) \cdot (x\equiv m y)$, 
defines $m/n$ in the variable $x$.
\end{proof}

We now establish that, under suitable axioms, variables can play the role of rational truth constants.
Let $Q = \{q_{m,n} \mid m,n\in N, m\leq n, n>0\}$ be a set of variables. 
We write $q_{m/n}$ instead of $q_{m,n}$. 
Define a theory $T_Q$ in {\L}ukasiewicz logic $\mathrmL$, 
using (only) the variables $Q$.
The theory $T_Q$ consists of:
\begin{itemize}
\item axiom $q_{0/n}\equiv \overline{0}$  for each $n > 0$;
\item axiom $q_{1/1} \equiv \overline{1}$;
\item axiom $q_{1/n} \equiv (\neg q_{1/n})^{n-1}$ for each $n\geq 2$;
\item axiom $q_{m/n} \equiv m q_{1/n}$ for each $m\leq n,n\geq 2$.
\end{itemize}
It follows from Corollary \ref{corollarymn} that 
$T_Q$ has exactly one model over $\standardL$, namely, 
the one where each $q_{m/n}$ is interpreted by $m/n$.

Under the axiom defining $1/n$, in the language containing the corresponding $q$-variable,
the value $m/n$ becomes \emph{term definable}
as $m q_{1/n}$; we might use this term directly, instead of introducing a new variable.
It is only introduced for uniformity of presentation.

\begin{lemma}
(Over {\L}ukasiewicz logic,) $T_Q$ proves the following:
\begin{itemize} 
\item each valid bookkeeping formula in the sense of $q$-variables; e.g., \\
$ q_{m/n} \cdot q_{k/l} \equiv q_{(m/n)\cdot(k/l)}$, and analogously for any other bookkeeping formula usually considered;
\item $q_{m/n} \equiv q_{m'/n'}$ for $m/n = m'/n'$;
\item $q_{m/n}\equiv \overline{1}$ for $m=n$.
\end{itemize}
\end{lemma}

\begin{proof}
We show the first statement; the rest is analogous. 
Consider a slightly stronger statement: let $\phi$ be a particular bookkeeping statement and let
$T_0\subseteq T_Q$ consist of definitions of only those rationals that are used in $\phi$ 
and any auxiliary rationals needed to define them; then $T_0$ is finite.
We have $T_0 \models_\standardL q_{m/n} \cdot q_{k/l} \equiv q_{(m/n)\cdot(k/l)}$ by definition of the $q$-variables.
The provability claim follows from finite strong standard completeness of $\mathrmL$.
\end{proof}


On an analogous note  (and with an analogous proof), one might substitute any other family of terms
that have been proved to enable implicit definability of the rationals (with a suitable indexing of variables): 
they will be provable from our axioms (and vice versa, each of the our axioms will be provable from them)
as a consequence of finite strong standard completeness.

The axioms just presented are redundant. It is in fact sufficient to define $1/n$ for \emph{each prime} $n$. 
Then not only $m/n$ becomes term definable for each $m\leq n$, but also, for $n,n'$ two primes, 
 $1/{(k n n')}$ is term definable for each $k$, 
because the subalgebra generated by $1/n$ and $1/n'$ in $\standardL$
contains the value $1/{(k n n')}$ and the algebra $\finiteMV{k n n'}$ generated by it; 
this entails that all of the elements are term definable (see \cite{DiNola-Leustean:HandbookMValgebras} and 
\cite{Torrens:CyclicElements}).

We define a translation $\star$ of formulas of RPL into  formulas of $\mathrmL$;
note that we assume RPL uses a countably infinite set $X$ of variables (we refer to them as $x$-variables).
For each variable $x_i$ of RPL, $x_i^\star$ is $x_i$,
for each constant (other than $\overline{0}$ and $\overline{1}$), $(m/n)^\star$  is $q_{m/n}$. 
Extend $\star$ to formulas as commuting with all connectives.
If $T$ is a set of RPL-formulas,
then $T^\star$ is the set of $\star$-translations thereof.
observe that $\star$ is a bijection between rational constants of RPL (different from $\overline{0}$ and $\overline{1}$) 
and $q$-variables of $\mathrmL$.
 
\begin{corollary}\footnote{Cf. \cite{Hajek:1998}, Lemma 3.3.13(2)} 
\label{interpret1}
Let $T$ be a theory and $\varphi$ a formula in the language of $\logic{RPL}$.
We have $T\vdash_{\logic{RPL}}\varphi$ if and only if $T^\star \cup T_Q \vdash_\mathrmL \varphi^\star$. 
\end{corollary} 
 
\begin{proof}
$\star$-translations of bookkeeping axioms are provable in $T_Q$. On the other hand, 
the $T_Q$-axioms become theorems of RPL under de-starring.
\end{proof}

As a special case ($T$ empty), the theory $T_Q$ proves (translations of) all theorems or RPL.

Coming back to bookkeeping formulas, let us see how they might provide an alternative implicit definition of the rationals.
We limit the presentation to a sketchy example, leaving the rest to the interested reader.
Working in a language with $\cdot$, $\to$ and $\overline{0}$ as basic connectives, we shall show how to define rationals with a denominator $3$. The following are some (not all) bookkeeping formulas for
variables indexed by such rationals:
\begin{align*}
x_{2/3} \cdot x_{2/3} &\equiv x_{1/3}\\
x_{1/3} \cdot x_{2/3} &\equiv x_0\\
x_{1/3} \cdot x_{1/3} &\equiv x_0\\
x_{1/3} \to x_0 &\equiv x_{2/3}
\end{align*}
Under the additional information that $x_0\equiv \overline{0}$, the last axiom reads $\neg x_{1/3} \equiv x_{2/3}$, which, combined with the first one, gives
$(\neg x_{1/3})^2 \equiv x_{1/3}$; the latter implies $v(x_{1/3})=1/3$ in $\standardL$ for all models $v$ of the four 
bookkeeping formulas above,
appealing to Corollary \ref{lemma1n}. 
Observations to be made are (a) the four formulas entail (semantically, and hence also syntactically) an instance of $x\equiv(\neg x)^{n-1}$ for $n=3$, and hence define $1/3$ in the variable $x_{1/3}$; (b)  the second and third axiom are redundant;
they follow (semantically, and hence syntactically) from the first and the fourth axiom.
Analogously, one might find redundancies in bookkeeping for other denominators, and hence also for combining different prime denominators,
for example. We do not go into detail.

Thus, bookkeeping formulas are sufficient (and more than that) for implicit definition of all the rationals in $\standardL$.

\subsection{Deductive Beth property}
\label{subsection:beth}

To formulate Beth property, we recall the definition of the two properties (of explicit and implicit definability)
as usually considered, i.e., referring to \emph{variables} rather than elements of a specific semantics. 
Within the scope of this subsection we shall use the notions as given in Definition \ref{beth_definitions}.

There are at least two ways to render the property in {\L}ukasiewicz logic 
(see the discussion in \cite{Montagna:Interpolation}); what is studied here is the \emph{deductive}
version of Beth property. 

\begin{definition}
\label{beth_definitions}
Let $\logic{L}$ be a logic and $\phi(x,\tilde{z})$ a term in the language of\/ $\logic{L}$. 
\begin{itemize}
\item  $\phi$ explicitly defines $x$ in $\logic{L}$ if and only if there is a term $\delta(\tilde{z})$ (in the $\tilde{z}$-variables only)
such that $\phi(x,\tilde{z}) \vdash_{\logic{L}} x \equiv \delta(\tilde{z})$ for each $x$ and $\tilde{z}$
\item  $\phi$ implicitly defines $x$ in $\logic{L}$ if and only if 
$\phi(x,\tilde{z}), \phi(x',\tilde{z}) \vdash_{\logic{L}} x \equiv x'$  for each $x$, $x'$, and $\tilde{z}$
\end{itemize}
\end{definition}

Our earlier (semantic) definitions of term- and implicit definability in $\standardL$ 
nod to the definitions (for variables) above.
In particular, any equation, or finite system thereof, that has a unique solution in a variable $x$ in $\standardL$,
translates in the obvious way into a propositional equivalence (a term) that implicitly defines $x$ in the logic $\mathrmL$.
This is a consequence of the finite strong standard completeness theorem of $\mathrmL$.

\begin{lemma} The term $x\equiv(\neg x)^{n-1}$ implicitly defines $x$ in $\mathrmL$.
\end{lemma}

We say that a logic has the deductive Beth property if, whenever a term $\phi$ implicitly defines $x$, then
the term also explicitly defines $x$.

\begin{corollary}
{\L}ukasiewicz propositional logic does not have the deductive Beth property.
\end{corollary}

\begin{proof}
We have that $x\equiv(\neg x)^{n-1}$ implicitly defines $x$. In order for this formula 
to explicitly define $x$, an MV-term needs to exist that is provably equivalent to $x$ under $x\equiv(\neg x)^{n-1}$; 
so in particular, in $\standardL$, the term would have to take the constant value $1/n$.
There is no such MV-term; therefore, $x\equiv(\neg x)^{n-1}$ does not explicitly define $x$ in $\mathrmL$.
\end{proof}

This result for {\L}ukasiewicz logic was given in \cite{Hoogland:Thesis}, 
addressing the topic in a comprehensive way over a large landscape of logics. See \cite{Montagna:Interpolation} 
for the more specific area of logics extending H\'ajek's Basic Fuzzy Logic $\BL$.


\subsection{Defining the irrationals}
\label{subsection:irrationals}

\begin{lemma}
An irrational number $a\in [0,1]$ is not definable by an MV-term in $\standardL$.
\end{lemma}

\begin{proof}
If an MV-term $\varphi(x_1,\dots,x_n)$ is satisfiable in $\standardL$, 
 there is a rational $n$-tuple $\langle r_1,\dots,r_n\rangle$ that satisfies $\varphi$. 
Let $a$ be an irrational. If $\varphi(x_1,\dots,x_n)$ is satisfiable in $\standardL$ and $1\leq i\leq n$, 
then there is an evaluation $v$ in $\standardL$ s.t.~$v(\varphi)=1$ and $v(x_i)\not=a$.
\end{proof}

Nevertheless, if one compromises on the requirement that implicit definitions are finite objects (i.e., given by a term) 
 and if one shifts from finitary to infinitary {\L}ukasiewicz logic $\infL$, 
one can implicitly define each irrational, as we now  show. 

Our simple construction illustrates the technique, working with only two values.
We leave any further elaboration to the reader. We do not discuss or advocate here
the necessity of introducing \emph{all} irrational values as constants.

We shall extend the theory $T_Q$. 
Consider irrationals $a,b\in [0,1]$ s.t. $a\cdot a=b$. 
Irrationals are cuts on the rationals, and the latter have already been defined by the axioms of $T_Q$.

Let
\begin{align*}
A_1 &= \{q\in \mathds{Q}\cap [0,1] \mid q < a \} \\ 
A_2 &= \{q\in \mathds{Q}\cap [0,1] \mid a < q \} 
\end{align*}
and analogously $B_1,B_2$ for b.
Introduce fresh variables (i.e., not occurring in $T_Q$) $i_a, i_b$. Let 
$$T_{Q,a} = T_Q \cup \{q_{m/n} \to i_a \mid m/n \in A_1\} \cup \{ i_a \to q_{m/n}  \mid m/n \in A_2\}$$
Analogously, define $T_{Q,b}$ for $b$.

\begin{lemma}
In any standard model of $T_{Q,a}$,
the variable $i_a$ has the value $a$.
\end{lemma}

\begin{proof}
Since $T_Q$ only admits canonical interpretations of the $q$-variables in the standard MV-algebra,
the cut $(A_1,A_2)$, captured by the axioms of $T_{Q,a}$, determines the valuation of $i_a$.
\end{proof}

Let $T_{Q,a,b}$ be the theory whose axioms are the union of axioms for $T_{Q,a}$ and $T_{Q,b}$.

\begin{corollary}
$T_{Q,a,b} \vdash_{\infL} i_a \cdot i_a \equiv i_b$.
\end{corollary}

\begin{proof}
Since the values of $i_a$ and $i_b$ are fixed in $\standardL$ by $T_{Q,a}$ and $T_{Q,b}$,
we have \\ $T_{Q,a,b} \models_{\standardL} i_a \cdot i_a \equiv i_b$.
Then \\ $ T_{Q,a,b} \vdash_{\mathrmL_\infty} i_a \cdot i_a \equiv i_b$ by strong standard completeness 
of $\mathrmL_\infty$. 
\end{proof}

As in the case of constants for rationals, one can introduce irrational constants incrementally (as above).
It is known that, if $a$ is an irrational in $\standardL$, then the subalgebra generated by $a$ is dense (\cite{Gaitan:SimpleOneGeneratedBCK});
as in the case of rationals, when just one value is defined by suitable axioms, other values become term definable 
within such a theory with an expanded language.

\subsection{Completeness theorems}

Finally, let us remark on completeness theorems in this new setting. There is nothing to add for general or standard finite strong completeness. 
However, one might wonder how to spell Pavelka completeness result.
Let $T$ be a set of formulas and $\varphi$ a formula of $\mathrmL$ (possibly with some $q$-variables).
We define
$$ ||\varphi||_{T\cup T_Q} = \inf \{ v(\varphi) \mid v \mbox{ model of }T\cup T_Q\}$$
and\footnote{Cf.~also \cite{Hajek-Paris-Shepherdson:Pavelka}, Theorem 3.1 and above.}
$$|\varphi|_{T\cup T_Q} = \sup\{m/n \mid T\cup T_Q\vdash_\mathrmL q_{m/n}\to \varphi \}$$
The definition rests on  a very convenient indexing of the $q$-variables. 
The same convention occurs in the usual spelling of the provability degree in RPL:
in particular, one can perform computations and take suprema on the names of the constants. 
Both definition rest on the link between the name of the constant and its intended value; 
usually the language is chosen in such a way that this link is immediate.

\section{Recognition and complexity}
\label{section:complexity}

To provide a sample application of the interpretation result,  
let us look at the complexity of provability from a finite theory in propositional $\RPL$;
in view of the finite strong standard (canonical) completeness theorem for RPL, 
this is just the finite consequence relation in $\standardL$ (for the propositional language with rational constants).
Recall the problem: given a finite theory $T$ and a formula $\phi$ of $\RPL$, does $T\vdash_{\RPL}\phi$?
For an empty $T$, this problem is just theoremhood in $\RPL$.
Without wishing to discuss implementation details, we remark that 
each rational constant is represented  as a pair of natural numbers given in binary.

A dual form of the conservativity result gives that provability without constants is a language fragment of provability with constants; for finite theories, the former is $\coNP$-complete, so the latter is $\coNP$-hard.
In \cite{Hajek:ComplexityRational}, 
H\'ajek argues $\coNP$-containment for (theoremhood and) provability from finite theories in $\RPL$  
by looking at the standard semantics 
and asking the reader to verify that the decision method used in \cite{Hajek:1998} for {\L}ukasiewicz logic,
which is obtained as a polynomial-time reduction to mixed integer programming, 
works even under the presence of rational constants in the formulas.
Of course, the extension with rational constants is quite natural for the mixed integer programming problem.

It is also unnecessary. The statement of Corollary \ref{interpret1} 
provides a method to argue the result without re-examining the reduction to mixed integer programming:
one simply uses the $\star$-translation and appeals to a finite subtheory of $T_Q$ to provide a canonical interpretation for the new variables;
in particular, for each constant $m/n$, it is necessary to include the defining axiom for $q_{1/n}$ and $q_{m/n}$ in the finite subtheory.
Next to that, the only complexity result needed is that of $\coNP$-completeness of finite consequence relation in $\logic{\L}$.

A small glitch is that the interpretation provided by $\star$ and (the finite subtheory of) $T_Q$ cannot be used in the form given above  because they do not yield a polynomial translation:
for a given $n$, the innocent-looking abbreviation $\phi^{n-1}$  in fact stands for a term of exponential size in $|n|$
(for $n$ represented in binary).
This problem can be rectified in the usual manner, using new variables for intermediate powers;
in the lemma below, powers of MV-terms and the product of MV-terms denoted by $\Pi$ 
 pertain to the multiplication symbol $\cdot$ of the language of MV-algebras.

\begin{lemma}
\label{lm_poly_traslation}
For $n\in N$, $n\geq 2$, take the binary representation of $n-1$, i.e., let $n-1=\Sigma_{i=0}^m p_i 2^i$ 
with $p_i\in\{0,1\}$ and $p_m=1$. 
Let $I=\{i\mid p_i=1\}$. 
In $\standardL$, the system of equations
\begin{align*}
y_0 &= \neg x  \\
y_1 &= y_0^2  \\
y_2 &= y_1^2  \\
&\mathrel{\makebox[\widthof{=}]{\vdots}} \\
y_m &= y_{m-1}^2\\
x &= \Pi_{i \in I} y_i
\end{align*}
has a unique solution, assigning the value $1/n$ to $x$.
\end{lemma}

\begin{proof}
The system implies $x=(\neg x)^{n-1}$. 
This equation has a unique solution $x=1/n$ in $\standardL$ (\cite{Hanikova-Savicky:SAT}, Lemma 6.3).
On the other hand, the assignment $x=1/n$ determines the values assigned to all the $y$-variables.
\end{proof}

In what follows, the $y$-variables from the above lemma are referred to as ``auxiliary variables''. 

\begin{lemma}
\label{lm_poly_2}
\begin{enumerate}
\item For $n,m,I$ as above, 
the system of formulas $$\{ y_0\equiv \neg q_{1/n}, \bigwedge_{i=1}^m (y_i = y_{i-1}^2),\,  q_{1/n} = \Pi_{i \in I} y_i\}$$
defines $1/n$ in the variable $q_{1/n}$ in $\standardL$. 
\item The size of the system is polynomial in the size of $n$.
\item Using an analogous argument, one can polynomially define $m/n$.
\end{enumerate}
\end{lemma}

\begin{theorem}
Provability from finite theories in $\RPL$ is polynomially reducible to provability from finite theories in $\mathrmL$.
\end{theorem}

\begin{proof}
Let a finite consecution $(T,\phi)$ in the language of $\RPL$ be given.
Let $\{ m_1/n_1,\dots m_k/n_k \}$ be all the rational constants in $T,q$, 
and let all variables therein be among $\{x_1,\dots, x_l\}$.
For $i=1,\dots,k$ let $T_i$ be the set of formulas from Lemma \ref{lm_poly_2} (1) and (3), using a fresh pool of auxiliary variables 
to define each of $1/n_i$ and  $m_i/n_i$.  
Let $T_Q^{fin} = \bigcup_{i=1}^k T_i$, and let $T^\star$ and $\phi^\star$ 
result from $T$ and $\phi$ by replacing its constants with the respective $q$-variables.
Then $T\vdash_{\logic{RPL}}\varphi$ if and only if $T^\star \cup T_Q^{fin}  \vdash_\mathrmL \varphi^\star$ as in Theorem \ref{interpret1};
moreover, the size of $ T_Q^{fin}$ (and also $T^\star$ and $\varphi^\star$) is polynomial in $|T|+|\phi|$. 
\end{proof}

\begin{corollary}
Provability from finite theories in $\RPL$ is in $\coNP$.
\end{corollary}

The construction of the polynomial translation is not less laborious than the inclusion of the rational constants in the mixed integer programming problem; 
one cannot argue that the (coNP-containment) result became simpler by omitting the constants.
 
However, what the construction shows is that problems such as the computational complexity of RPL are in fact problems about 
$\mathrm L$ (using the translation).
In other words, there was no question one could ask about complexity of RPL in the first place,
other than those already settled by complexity results for $\mathrmL$.

\bigbreak

Corollary \ref{corollarymn} provides a way of defining each of the rationals in {\L}ukasiewicz logic
in the sense of Definition \ref{def_definability}. 
This is just one among many variants of the implicit definability result.
One can observe that the construction rests on a parametrized formula where 
by varying the parameter one easily gets the 
definition of the intended constant for each $m/n$.

On the other hand, consider a randomly chosen formula. What, if anything, does it define? 
We formulate this question as a decision problem, and show that it is hard.

\begin{problem}
Given an MV-term of $n$ variables $\varphi(x_1,\dots,x_n)$, an integer $1\leq i\leq n$, and a rational number $a$,
 determine whether $\varphi$ defines $a$ in $x_i$ (in $\standardL$). 
\end{problem}

 The decision problem on the domain of triples $\langle \varphi,i,a\rangle$ as above,
 consisting of exactly those where $\phi$ defines $a$ in $x_i$ in the standard MV-algebra,
 is referred to as the $\DEF$ problem.

We seek to estimate the complexity of the problem.
The estimate will in part be an artefact of our definition. 
Namely, we have stipulated that each of our defining formulas be satisfiable (in $\standardL$); 
otherwise, if only the second item in Definition \ref{def_definability} were used, each 
unsatisfiable formula would implicitly define every rational (and irrational) in a trivial way.
The following lemma reflects our choice.

\begin{lemma}
\label{lemma_def_NPhard}
$\DEF$ is $\NP$-hard.
\end{lemma}

\begin{proof}
We reduce $\SAT$ to $\DEF$ (both notions relate to $\standardL$).

The reduction function $f$ assigns, to a given formula $\phi(x_1,\dots,x_n)$, the triple 
$$\langle \phi \cdot (x_{n+1}\equiv\neg x_{n+1}), n+1, 1/2\rangle$$ 
We show that $\phi \in \SAT$ if and only if $f(\phi)\in \DEF$.

Let $\phi$ be satisfiable; then so is $\phi \cdot (x_{n+1}\equiv\neg x_{n+1})$, and moreover, 
if the latter is satisfied by a $v$, then
$v(x_{n+1})=1/2$, because $1/2$ is the only solution to the equation 
$x=\neg x$ (note $x_{n+1}$ does not occur in $\phi$).

On the other hand, if $\phi$ is unsatisfiable, then so is $\varphi \cdot (x_{n+1}\equiv\neg x_{n+1})$.
\end{proof}


\begin{theorem}
\label{th_def_coNPhard}
$\DEF$ is $\coNP$-hard.
\end{theorem}

\begin{proof}
For a given $\langle \varphi,i,a\rangle$ as above, let us consider the condition
\begin{equation}\exists v_{\standardL} ( v(\phi)=1 \et v(x_i)\not= a)\tag{D}\end{equation}

This condition negates the second condition in Definition \ref{def_definability}. 
It is in $\NP$ (because of small witnesses); to show it is $\NP$-hard, it is enough to consider a variant 
of the reduction from Lemma \ref{lemma_def_NPhard}: to find out whether $\varphi(x_1,\dots,x_n) \in \SAT$,
ask the algorithm for D about $\langle \varphi \cdot (x_{n+1}\equiv\neg x_{n+1}), n+1, 1/4\rangle$.
If $\phi$ is satisfiable, then D is satisfied on $f(\phi)$; if not, then neither is $\varphi \cdot (x_{n+1}\equiv\neg x_{n+1})$, 
so D fails. So D is $\NP$-complete.

Now we reduce $\bar D$ (the complement of D),  i.e., the set $\langle \varphi,i,a\rangle$ s.t.
$v(\varphi)=1$ implies $v(x_i)=a$, to $\DEF$ (i.e., to $\bar D$ with the additional condition of satisfiability for $\varphi$).

Let $\langle \varphi(\tilde{x}),i,a\rangle$ be an instance. 
Let $\psi(x_i,\tilde{z})$ be a formula defining $a$ in $x_i$  (as earlier in this paper).
Note that \\
(i) $\psi$ is satisfiable, and for any satisfying evaluation $v$, we have $v(x_i)=a$;\\
(ii) $\psi$ can be chosen so that its size is polynomial in the size of the given instance.\\
Let $f(\langle \varphi,i,a\rangle) = \langle (\varphi \vee \psi) (\tilde{x},\tilde{z}) ,i,a\rangle$. 
We claim that $\langle \varphi,i,a\rangle$ satisfies (D) if and only if $f(\langle \varphi,i,a\rangle) \in \DEF$.

If $\langle \varphi,i,a\rangle$ satisfies $\bar D$, then $\varphi\lor\psi$ is satisfiable (because $\psi$ is), 
and satisfies the condition of $\bar D$, because both $\varphi$ and $\psi$ satisfy it. So $f(\langle \varphi,i,a\rangle)\in \DEF$.

If $\langle \varphi,i,a\rangle$ satisfies (D), then there is a $v$ s.t. $v(\varphi)=1$ and $v(x_i)\not=a$. 
This is also true about $\varphi\lor\psi$.
 \end{proof}

 (The proof of) Theorem \ref{th_def_coNPhard} answers the complexity question for a modification 
 of Definition \ref{def_definability} omitting the satisfiability requirement. Such a problem is $\coNP$-complete.

\section{More on RPL and {\L}}

As stated already, $\mathrmL$ is a syntactic fragment of $\RPL$:
well-formed formulas of $\mathrmL$ are exactly those well formed formulas of $\RPL$
that do not contain any constants other than $\overline{0}$ and $\overline{1}$.
Dually, $\RPL$ is a conservative extension of $\mathrmL$.

The conservative extension statement, however, does not quite capture the tight relation between $\RPL$ and $\mathrmL$.\footnote{This is
also reflected in  \cite[{\textsection}3]{Hajek-Paris-Shepherdson:Pavelka}:``Lemma 2.3 shows that $\RPLpred$ is a very conservative extension
indeed of $\Lpred$. There is a sense in which even its new formulae don't express anything which can't be expressed by old formulae.''}
To appreciate that, consider for example that Peano arithmetic is a conservative extension of Presburger arithmetic.
But Peano arithmetic does \emph{not} stand to Presburger arithmetic
as $\RPL$ stands to $\mathrmL$.

Rather, what we have in $\RPL$ is an axiomatic expansion of the language as exemplified by extending the theory of groups,
presented with the binary group operation only, with a new constant for the neutral element.
This is an extension by definition (so it is conservative) and because we can prove existence and unicity
of the solution to $x\circ x=x$, the new constant for the neutral element can be eliminated.
In the case of $\RPL$, \cite[Theorem 2.6]{Hajek-Paris-Shepherdson:Pavelka} gives the exact sense in which
constants can be eliminated, as mentioned in Section \ref{section:intro} above.

Not so much the conservativity result, but predominantly
the faithful interpretation result suggests that the logics stand too close to each other to be even 
considered as two genuine systems, except for some very theoretical considerations (such as the set of term-definable
functions in the standard semantics, for example); for practical applications, however, it is a matter of convenience
which of the two logics is used.

\section{On precision and vagueness}
\label{section:vagueness}

Within the framework of  {\L}ukasiewicz logic, with or without constants in the language,
one can be surgically precise about truth values ascribed to particular propositions.
We have seen that, under a simple theory such as $T_Q$, 
one can pinpoint the truth value of a proposition $p$ to any rational number by positing
$p\equiv q_{m/n}$. 
There is no talk about valuations here: $T_Q$ and $p\equiv q_{m/n}$ MV-terms or sets thereof.\footnote{Admittedly, $T_Q$ imposes its own semantics: it does not have a model over algebras that do not contain a copy of the rationals.}
Given that the rationals are dense within $[0,1]$, this level of precision seems sufficient for practical purposes,
which presuppose finiteness. Should irrational values be needed, however (thus forfeiting finiteness), 
we have shown how to obtain them switching the underlying logic to the infinitary logic $\infL$.

In recent papers concerning the modelling of vague predicates with fuzzy logic, the predicate \emph{tall} is considered
as an example amenable to modelling by the apparatus of formal fuzzy logic. It is so amenable because it is a vague way of referring to an easily measurable quantity, namely \emph{height}, which is not vague, 
but is measured on a linear scale.

The claim here is not that {\L}ukasiewicz logic or its unspecified theories \emph{prove} $p$ to have the truth value $m/n$ 
or any other value for that matter (unless $p$ turns out to be a theorem or a contradiction of {\L}ukasiewicz logic, of course).
Specific theories may indeed prove $p\equiv q_{m/n}$, 
but then (trivially) the axioms of such theories need to be \emph{at least as strong} 
as $p\equiv q_{m/n}$.
{\L}ukasiewicz logic however, as a formal system, does not commit to such ad hoc stipulations, 
but merely makes it possible to \emph{express} them, and to enable deduction on them.

One possible rephrasing of the above is that, right side of the turnstile, 
we know rather little of how our ``truth constants''  came to be; 
bar some notational conventions (helping us to distinguish between constants and variables), 
we are presented with a set of syntactic objects 
that obey some bookkeeping statements.

Detailed  accounts \cite{Marra:PrecisionVagueness} and \cite{Behounek:WhichSenseFuzzy},
echoing also earlier works of H\'ajek, explain that ({\L}ukasiewicz or other) fuzzy logic does not insist on 
assigning specific real values to propositions, but rather, broadly speaking, studies consequence on, and reasoning with, 
propositions that in principle admit  these truth values. These accounts are offered in response to some attempts 
at refutation of fuzzy logic on the ground of the artificial precision problem that points out
the lack of incentives for preferring one intermediate truth value over another one for particular propositions.

We have argued that precision can be neatly captured by the syntax of {\L}ukasiewicz logic.  
All of that is happening \emph{left of the turnstile}. We have also reminded ourselves of the simple fact that
it is nontrivial to recognize what precisely an arbitrary theory implicitly defines.
One might, therefore, understand the account presented here as a supportive argument for
the thesis that precision happens in fuzzy logic, and if it happens, it must happen within the assumptions 
before it can happen within the conclusions, which seems to be one of the main points of the above papers.

\section{Conclusion}

Thanks to finite strong standard completeness of {\L}ukasiewicz logic, 
our semantic rendering of implicit definability blends naturally with the more commonly given implicit definability notions
that refer to variables (as when introducing Beth property, see \cite{Hoogland:Thesis,Montagna:Interpolation})
or connectives (as in \cite{Caicedo:ImplicitConnectivesAlgebraizable,Caicedo:ImplicitOperationsMV}), i.e., syntactic objects. 

The findings of Section \ref{section:definability} and \ref{section:complexity} seem to confirm 
the ambivalent status of constants in {\L}ukasiewicz logic:
they are useful but dispensable abbreviations, with statements of {\L}ukasiewicz logic with constants
being expressible in {\L}ukasiewicz logic without any constants but those that are inherent to it.
It may well be that keeping the constants is ``more elegant'', as asserted by H\'ajek in \cite{Hajek:1998},
or that ``\dots even for partial truth, Rational Pavelka logic deals with exactly the same logic as {\L}ukasiewicz logic---but in a very much
more convenient way'', as \cite{Hajek-Paris-Shepherdson:Pavelka} remarks.
We do not attempt to argue for or against these views.
What is argued here, however, is that  metamathematical statements about Rational Pavelka logic  
naturally translate to statements about (certain theories over) {\L}ukasiewicz logic without any added constants.
In particular, it is not the case that {\L}ukasiewicz logic with truth constants is 
richer, more expressive, or more general than {\L}ukasiewicz logic without constants.

\newpage
\noindent
{\bf Acknowledgements.} The author was supported by CE-ITI and GA\v CR 
under the grant number GBP202/12/G061, and by the long-term strategic development financing of the Institute of Computer Science RVO:67985807.

\bibliographystyle{plain}

\end{document}